\documentclass{article}
\usepackage{emulateapj}
\usepackage{epsfig}
\usepackage{flushrt}
\usepackage{apjfonts}

\newcommand{\bvfreq}{Brunt-V\"ais\"al\"a frequency}
\newcommand{\etal}{et al.}
\newcommand{\subs}[1]{_{\rm #1}}
\renewenvironment{figure}{\begin{figure*} \centering}{\end{figure*}}

\begin{document}

\submitted{Received 1998 October 23; accepted 1999 May 25}

\title{Evolutionary Calculations of Phase Separation in Crystallizing
White Dwarf Stars}

\author{
M. H. Montgomery,\altaffilmark{1,2}
E. W. Klumpe,\altaffilmark{1}
D. E. Winget,\altaffilmark{1} \&
M. A. Wood\altaffilmark{3}
}

\altaffiltext{1}{McDonald Observatory and Department of Astronomy,
The University of Texas, Austin, TX 78712 - USA.}
\altaffiltext{2}{Institut f\"ur Astronomie, Universit\"at Wien,
T\"urkenschanzstra\ss e 17, A-1180 Wien, Austria.}
\altaffiltext{3}{Department of Physics and Space Sciences and SARA
Observatory, Florida Institute of Technology, Melbourne, FL 32901-6988 USA}

\begin{abstract}
We present an exploration of the significance of Carbon/Oxygen phase
separation in white dwarf stars in the context of self-consistent
evolutionary calculations.  Because phase separation can potentially
increase the calculated ages of the oldest white dwarfs, it can affect
the age of the Galactic disk as derived from the downturn in the white
dwarf luminosity function.  We find that the largest possible increase
in ages due to phase separation is $\sim $1.5 Gyr, with a most likely
value of approximately 0.6 Gyr, depending on the parameters of our white
dwarf models.

The most important factors influencing the size of this delay are the
total stellar mass, the initial composition profile, and the phase diagram
assumed for crystallization. We find a maximum age delay in models with
masses of $\sim 0.6 M_{\odot}$, which is near the peak in the observed
white dwarf mass distribution.  In addition, we note that the prescription
which we have adopted for the mixing during crystallization provides
an upper bound for the efficiency of this process, and hence a maximum
for the age delays. More realistic treatments of the mixing process
may reduce the size of the effect.  We find that varying the opacities
(via the metallicity) has little effect on the calculated age delays.

In the context of Galactic evolution, age estimates for the oldest
Galactic globular clusters range from 11.5 to 16 Gyr, and depend on
a variety of parameters. In addition, a 4 to 6 Gyr delay is expected
between the formation of the globular clusters and that of the Galactic
thin disk, while the observed white dwarf luminosity function gives an age
estimate for the thin disk of $9.5^{+1.1}_{-0.8}$ Gyr, without including
the effect of phase separation. Using the above numbers, we see that
phase separation could add between 0 to 3 Gyr to the white dwarf ages
and still be consistent with the overall picture of Galaxy formation.
Our calculated maximum value of $\lesssim$ 1.5 Gyr fits within these
bounds, as does our best guess value of $\sim$~0.6 Gyr.

\end{abstract}


\section{Astrophysical Context}

The phenomenon of phase separation and crystallization exists within
the larger context of white dwarf cooling. Since the time of Mestel's
original treatment (\cite{Mestel52} 1952), much work has been done, both
to improve the input physics of the models and to make more complete
observations of the white dwarf luminosity function (WDLF). In 1987,
\cite{Winget87} showed that the observed downturn in the WDLF could be
understood in terms of a finite age for the Galactic disk, and that the
WDLF could therefore in principle be used to determine an age for the
local Galactic disk.  Using the preliminary results from \cite{Liebert88}
(1988, hereafter LDM) for the observed WDLF, they obtained an age for
the local Galactic disk in the range 7--10 Gyr.  Since then, Wood has
made more detailed calculations using improved input physics, Galactic
evolution models, and WD parameters to constrain this age even further
\cite{Wood92} (1990, 1992, 1995). Historically, these developments
were foreshadowed by \cite{Schwarzschild58} (1958), \cite{Schmidt59}
(1959), and \cite{D'Antona78} (1978), all of whom considered white dwarf
evolution in a Galactic context.

Two observational surveys within the last ten years stand out in their
importance to the field. First, Liebert et al. (1988) produced a WDLF
containing 43 cool field WD's, which was the largest such sample size up to
that point in time. More recently, \cite{Oswalt96} (1996) produced a
WDLF of 50 cool WD's in wide binaries.  Using the models of \cite{Wood95},
the LDM sample yields an age for the Galactic Disk of $\sim 7.5 \pm 1$
Gyr (\cite{Wood95} 1995), while the \cite{Oswalt96} (1996) sample gives
an age of $9.5^{+1.1}_{-0.8}$ Gyr. Taking the error estimates at face
value, these results differ by $2 \sigma$. \cite{Wood98} (1998) conducted
Monte Carlo simulations and found that it is unlikely that both samples
are consistent with the same parent population. Further investigation 
will be needed to resolve the cause of this discrepancy.

In addition to the uncertainties in the observed WDLF, the way we treat
various physical processes in white dwarf interiors greatly affects the
ages that we derive for them. After the prediction in the early 1960's
that white dwarfs should undergo a phase transition and crystallize
as they cool (\cite{Abrikosov60} 1960; \cite{Kirzhnits60} 1960;
\cite{Salpeter61} 1961), \cite{Mestel67} (1967) and \cite{VanHorn68}
(1968) estimated that the associated release of latent heat during this
process would be large enough to delay the cooling of white dwarfs
significantly. \cite{Lamb75} (1975) included this energy release as
part of their evolutionary calculations of a 1 $M_{\odot}$ pure carbon
white dwarf.

\cite{Stevenson77} (1977) was the first to propose a phase separation
model that might affect white dwarf cooling times by providing an
additional source of energy analogous to the release of latent heat. This
model had a carbon core with trace amounts of iron. In a later model,
\cite{Stevenson80} (1980) suggested that a uniform mixture of carbon
and oxygen would become chemically differentiated as a result of the
crystallization process. Because such a redistribution of elements could
lower the binding (non-thermal) energy of the star, the change in energy
would be added to the thermal energy, and hence the luminosity, of the
star. This would increase the time for a white dwarf to cool to a given
luminosity, and would extend the apparent age of the Galactic disk as
derived from the WDLF.

Estimates of the amount by which the age of the local Galactic disk might
be extended have ranged from 0.5 Gyr to 6 Gyr (\cite{Mochkovitch83} 1983;
\cite{Barrat88} 1988; \cite{Garcia-Berro88} 1988; \cite{Chabrier93} 1993;
\cite{Segretain93} 1993; \cite{Hernanz94} 1994; \cite{Segretain94} 1994;
\cite{Isern97} 1997; \cite{Salaris97} 1997), although recent estimates
have been on the smaller end of this range, e.g., \cite{Salaris97} (1997)
calculate a delay of $\sim 1.0$ Gyr. Most of this spread in calculated age
delays comes from differences in the assumed phase diagram, although the
assumed C/O profile also has a large effect. 

In the context of Galactic evolution, age estimates for the oldest
Galactic globular clusters range from 13--16 Gyr (\cite{Pont98} 1998)
to 11.5 $\pm$ 1.3 Gyr (\cite{Chaboyer98} 1998), and depend on a variety
of parameters. In addition, a 4 to 6 Gyr delay is expected between the
formation of the globular clusters and that of the Galactic thin disk
(e.g., \cite{Burkert92} 1992; \cite{Chiappini97} 1997), while the observed
white dwarf luminosity function gives an age estimate for the thin disk
of $9.5^{+1.1}_{-0.8}$ Gyr (\cite{Oswalt96} 1996), without including the
effect of phase separation. Using the above numbers, we see that phase
separation could add anywhere from 0 to 3 Gyr to the white dwarf ages
and still be consistent with the overall picture of Galaxy formation.

In this paper, we examine the sensitivity of this calculated age delay
to the various physical assumptions by varying the initial C/O profile
of the white dwarf models, their total mass, and their H and He surface
layer masses. In addition, we examine the effect of using two different
published phase diagrams for the phase separation process, that of
\cite{Segretain93} (1993) and that of \cite{Ichimaru88} (1988). 

Our work improves upon previous calculations of the age delay in that
we use self-consistent evolutionary models. In particular, our models
use the modern OPAL opacities (\cite{Iglesias93} 1993) instead of the
older Cox-Stewart opacities, and we are able to treat self-consistently
the age delay as a function of total stellar mass, instead of using a
relation scaled by mass for the connection between the core temperature
and the surface luminosity. Finally, we are able to examine surface
layer masses suggested by more recent asteroseismological investigations
(\cite{Clemens93} 1993).

\section{The Physics of Phase separation}
\subsection{Chemical Redistribution}

Our present physical picture for the phenomenon of phase separation in
white dwarf stars is as follows.  As a white dwarf cools, it eventually
reaches a temperature when its central regions begin to crystallize. This
occurs when the thermal energy of the ions becomes much smaller than the
energy of the Coulomb interactions between neighboring ions. As a result,
the ions settle into lattice sites and lose the ability to move freely
in three dimensions.

If the white dwarf interior is initially a mixture of C and O, then
recent calculations indicate that the solid which crystallizes will
have a higher O content than the fluid from which it formed (Ichimaru
\etal\ 1988; \cite{Segretain93} 1993). Thus the crystallizing region
of the white dwarf becomes O-enhanced and the fluid layer overlying
this region becomes C-enhanced. Since the C is slightly less dense than
the O at a given pressure, this C-enhanced fluid layer is mixed via a
Rayleigh-Taylor instability (\cite{Mochkovitch83} 1983; \cite{Isern97}
1997) with the layers above, and C is transported outward from the center.
As the white dwarf continues to crystallize, the O-enhanced crystalline
core also continues to grow, with the net result that O is transported
inward in the white dwarf and C is transported outward. Thus, the chemical
composition profile after significant crystallization has occurred is
different from the profile before crystallization.

Just how different this profile is depends on the particular phase
diagram which is adopted for the process. In a ``spindle'' diagram,
the solid which forms always has an enhanced concentration of the
higher charge element (in this case oxygen), and the temperature of
crystallization of the mixture lies between that of the individual
elements. An ``azeotropic'' diagram differs from this in that there is a
range of concentrations for which crystallization takes place {\em below}
the temperature of crystallization of either of the pure elements. This
is somewhat analogous to the phenomenon of ``supercooling.'' Finally, a
``eutectic'' phase diagram is one in which there is a near total separation
of the higher and lower charged ions upon crystallization, resulting in
a segregation of the two chemical species.

Stevenson's original phase diagram (\cite{Stevenson80} 1980) was a
eutectic phase diagram with C and O being immiscible in the solid phase,
with the result that a pure O core would be formed in the models during
crystallization. Using a density functional approach, Barrat \etal\
(1988) calculated a phase diagram of spindle type.  In this case, the
solid which forms is a C/O alloy, but with the O content of the solid
enhanced relative to that of the fluid out of which it formed.

\begin{figure}
\epsfig{file=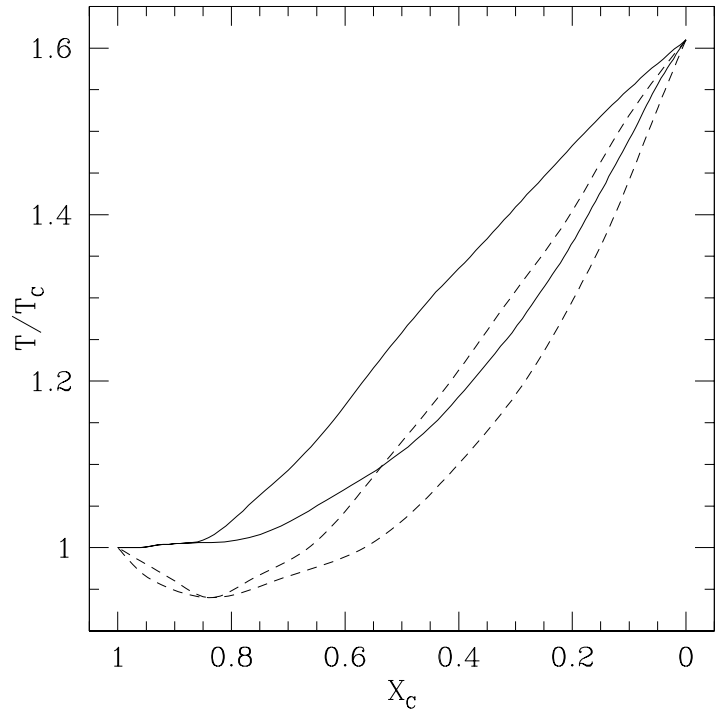,width=3.9in}
\caption{Phase diagrams for a C/O mixture as computed by Ichimaru \etal\
  (1988, dashed line) and \cite{Segretain93} (1993, solid line), where
  the vertical axis is in units of the crystallization temperature
  of C, $T_C$, and the horizontal axis is the C mass-fraction.  The solid
  line is of ``spindle'' type, while the dashed line is that of an
  ``azeotrope.'' The principal feature of the azeotrope is that there
  is a range of compositions for which the crystallization temperature
  of the mixture is less than that of either of the two constituents in
  the pure state.
  \label{fig1} }
\end{figure}

This problem was revisited by Ichimaru \etal\ (1988). They found that
Stevenson's initial prediction of a eutectic phase diagram was an
artifact of his use of the random-alloy mixing (RAM) model for the
internal energies in the solid phase. By comparison with Monte Carlo
simulations, they found that the linear mixing formula is more accurate
for the solid phase. They then used density-functional theory to derive
a phase diagram of azeotropic type, which is shown as the dashed line in
Figure \ref{fig1}. This diagram is similar to the spindle diagram,
with the exception that there is a range of compositions for which the
crystallization temperature is less than the crystallization temperature
for either of the pure compositions.

Most recently, \cite{Segretain93} (1993) used a density-functional
approach to derive phase diagrams for arbitrary binary-ionic mixtures,
as a function of $Z_1/Z_2$, where $Z_1$ and $Z_2$ are the nuclear charges
of the two chemical species. For C and O ($Z_1/Z_2=0.75$), they obtain
a phase diagram of spindle type, which is shown as the solid line in
Figure \ref{fig1}.

As shown in Figure \ref{fig1}, these diagrams of Ichimaru \etal\
and \cite{Segretain93} differ slightly in the composition changes during
crystallization, as well as in the temperatures at which crystallization
takes place. As a result, they produce different chemical composition
profiles after crystallization and different age delays.

We mention that the validity of the azeotropic phase diagram of Ichimaru
\etal\ (1988) has recently been called into question by \cite{DeWitt96}
(1996). DeWitt \etal\ find that the azeotropic diagram of Ichimaru \etal\
results from their use of the fitting function of \cite{Ogata87} (1987)
for the energies of the pure phases. The slight inaccuracies in this
fitting function lead to incorrect values of the free energy for the
pure phase, which in turn leads to a spurious departure from the linear
mixing rule in the liquid. With this proviso, we have elected to include
calculations based on the phase diagram of Ichimaru \etal\ for purposes
of comparison, in order to illustrate some of the uncertainties to which
these calculations can be subject.

\subsection{Energy Release}

Because the distribution of C and O within a model changes during phase
separation, the density profile changes as well. At a given pressure, O is
slightly denser than C. A model which has undergone phase separation
has more oxygen in its core, and thus a slightly larger concentration of
mass in its central regions. As a result, the phase-separated model is
more tightly bound gravitationally.

While it may be convenient to think of the energy which is released as
being due solely to the change in the gravitational potential energy
of the star, this is only part of the story. The relevant quantity is
actually the total binding energy of the star, $E\subs{bind}$, which
is the sum of all the nonthermal (structural) sources of energy. As such,
it acts as a potential energy for the configuration.
$E\subs{bind}$ can be written as
\begin{equation}
E\subs{bind} = E\subs{grav} + E\subs{deg} + E\subs{coul}
\end{equation}
where $E\subs{grav}, E\subs{deg},$ and $E\subs{coul}$ are the respective
energy contributions from gravitational interactions, kinetic energies
of the degenerate electrons, and Coulomb interactions among the different
charged particles (ions and electrons).

As phase separation occurs and the central regions become oxygen
enriched, the central density of the model increases. Thus,
$E\subs{grav}$ becomes more negative, as does $E\subs{coul}$.
$E\subs{deg}$, however, becomes more positive, since the Fermi energy
of the electrons increases with increasing density.  Summing these
contributions, we find that there is a net decrease in $E\subs{bind}$
for the models considered here. Due to conservation of energy, this
energy must be used to increase the thermal energy of the ions, which
are the only significant repository of thermal energy in the cores of our
white dwarf models.   This energy, then, is available to be radiated away 
and acts as an additional luminosity source.

The various contributions to the photon luminosity $L$ and the neutrino
luminosity $L_{\nu}$ of the white dwarf may be formally written as
(e.g., \cite{Isern97} 1997; \cite{Chabrier98} 1998)
\[
L + L_{\nu}  =  - \int_0^{M\subs{WD}} dm \left[ C_v \frac{dT}{dt} + \right. 
\]
\begin{equation}
\left.
T \left(\frac{\partial P}{\partial T} \right)_{V,X_O} \frac{dV}{dt} + 
\left(\frac{\partial u}{\partial X_O} \right)_{T,V} \frac{dX_O}{dt}
  \right]
\label{luminosity}
\end{equation}
where $V = 1/\rho$ is the specific volume, $X_O$ is the mass-fraction
of the heavier of the two chemical species (in this case, oxygen),
and $u$ is the internal energy per unit mass, which contains thermal,
electron degeneracy, and Coulomb contributions.  The first term in the
integrand on the righthand side of equation \ref{luminosity} is due to
the heat capacity of the core, which includes the release of latent heat
of crystallization, while the second term gives the contribution to the
luminosity due to volume changes, and is usually small in white dwarfs
since the pressure $P$ is only a weak function of the temperature. The
final term gives the luminosity due to the changing chemical composition
profiles within the white dwarf. This is the term we will study in our
numerical calculations.

As a check on the direct evolutionary calculations, we can estimate
the age delay produced by a given energy release. If we denote by $dE$
a small amount of energy which is released during the process of
phase separation, and if we assume that this energy is quickly
radiated, then we can calculate an estimated age delay $t_d$:
\begin{equation}
t_d = \int \frac{dE}{L}.
\label{td1}
\end{equation}
In the context of a sequence of evolutionary models, this integral is
operationally a sum, since a given model is computed at discrete
points in time, luminosity, etc. Furthermore, since the energy $\Delta
E_{i}$ is released {\em between} luminosities $L_{i-1}$ and $L_{i}$,
say, the average luminosity at which the energy is released is
approximately $(L_{i-1}+L_{i})/2$, so the discrete version of equation
\ref{td1} becomes
\begin{equation}
t_d = \sum_i \frac{\Delta E_i}{\case{1}{2}(L_{i-1}+L_{i})}.
\label{td2}
\end{equation}
We have used equation \ref{td2} as an alternate prescription to calculate
age delays. For the larger energy releases, $t_d$ computed in this way
agrees with the delay calculated from the self-consistent evolutionary
calculations, and for small energy releases it provides a better estimate
since the small energies can become masked in the numerical noise of the
evolutionary calculations.

\section{Numerics}

The basis for these calculations is WDEC, the White Dwarf Evolutionary
Code, as described in \cite{Lamb75} (1975), and in \cite{Wood90}
(1990). Our current version uses the updated OPAL opacity tables
(\cite{Iglesias93} 1993; \cite{Wood93} 1993). We use the additive volume
technique to treat the equation of state of the C/O mixture in the cores
of our models.

\subsection{The Melting Curve}

Our criterion for crystallization is given by the phase diagram which we
adopt, with the following caveat. Our equation of state (EOS) is based
on the Lamb EOS code (\cite{Lamb75} 1975), which has $\Gamma \simeq 160$
at crystallization. Here, $\Gamma \equiv Z^2 e^2/ \langle r \rangle
k_B T$ is the ratio of Coulomb energy between neighboring ions to each
ion's kinetic energy.  More recent calculations indicate that $\Gamma
\simeq 180$ (\cite{Ogata87} 1987). As a result, our values for the
crystallization temperature of C, $T\subs{C,xtal}$, are too high by a
factor of $\sim 180/160 = 1.125$.

To remedy this situation, we could simply adjust $T\subs{C,xtal}$ downward
accordingly, and we have done this for a few runs. This is inconvenient,
however, because it places us at the edge of our EOS tables which were
calculated with $\Gamma \simeq 160$.  Instead, we apply a correction
factor to our calculated age delays which takes into account the fact that
crystallization/phase separation occurs at lower central temperatures,
and therefore lower luminosities, than is calculated directly in our
models. This correction to the calculated age delays is typically of
order 25\%. For example, an age delay due to phase separation computed
with $\Gamma \simeq 160$ might be $\sim$ 1 Gyr, which with the more
physical value of $\Gamma \simeq 180$ would be $\sim$ 1.25 Gyr.  We find
that this procedure of computing the age delays based on the corrected
luminosity during crystallization to be accurate to within 1--2\%.

\subsection{Implementation in WDEC}

The calculation of the evolutionary sequences is ``quasi-static'' in the
sense that we compute a sequence of static models separated by finite
steps in time. Each static model represents the cooling white dwarf at a
different age and luminosity. We include the physics of phase separation
using the same approximation:  we assume that the timescale for any mixing
which occurs is short compared to the individual evolutionary timesteps
(see section \ref{consistency} of this paper; \cite{Mochkovitch83}
1983; \cite{Isern97} 1997), and we assume that the binding energy which
is released by this process can be modeled by some suitably chosen local
energy generation rate, $\epsilon\subs{ps}$ (e.g., \cite{Isern97} 1997).

The phase separation calculation may therefore be broken into three
sections. The first part involves obtaining the changing composition
profile as a function of the crystallized mass-fraction, while the
second part is the calculation of the cumulative energy released, also
as a function of the crystallized mass-fraction. The final part is the
calculation of the value of $\epsilon\subs{ps}$, which is the energy
locally deposited per unit mass per unit time. Our implementation of the
complete problem is self-consistent in that we let $\epsilon\subs{ps}$
vary as the compositional profile changes due to crystallization, as
WDEC iterates to a converged model.

The first part of the overall problem relates to the composition of
the crystallizing layers. Using the phase diagram of \cite{Segretain93}
(1993) or Ichimaru \etal\ (1988), we compute the final composition profile
of the model given the initial profile, before doing a full evolutionary
calculation.  This is possible because the composition of the crystals
which are forming is determined solely by the mass fractions of C and
O which are present in the fluid phase, and not by the temperature and
density of the medium (the temperature and density of course determine
{\em when} the fluid crystallizes, but given that it is crystallizing,
the chemical composition of the solid is determined {\em solely} by the
composition of the fluid). We are therefore required to compute only once,
at the onset of crystallization, the composition profile as a function of
the crystallized mass fraction.  At subsequent evolutionary times, we use
this relation and the current crystallized mass-fraction to interpolate
onto the composition grid, which is a computationally convenient procedure.

We take this same approach for the calculation of the energy released. At
the onset of crystallization, we calculate the total amount of energy
released as a function of $M\subs{xtal}/M\subs{\star}$, using the relation
\begin{equation} \delta E = \int_0^{M\subs{WD}} \left(\frac{\partial
u}{\partial X_O} \right)\subs{T,V} \delta X_O dm, \label{pscalc}
\end{equation} where $\delta E$ is the binding energy released by the
composition change $\delta X_O$. Since these changes in composition are
with respect to the pre-crystallization state, we are in effect holding
both the temperature and density profiles constant for all subsequent
phases of crystallization.  Holding the temperature profile constant is a
quite reasonable approximation, since the vast majority of the mass in the
the white dwarf model is strongly degenerate for the temperature range of
interest. Similarly, we expect the changes in the density profile to be
small ($\frac{\delta \rho}{\rho} \lesssim$ 1\%) even in the presence of
composition changes.  This is a consequence of the fact that the equations
of state for carbon and oxygen are very similar in the strongly degenerate
regime, i.e., $\mu\subs{e}$, the atomic mass per electron, is 2.0 for
both elements. This suggests that this approach would not necessarily
be as accurate for carbon and iron, since $\mu\subs{e} = 2.15$ for iron.

\begin{figure}
\epsfig{file=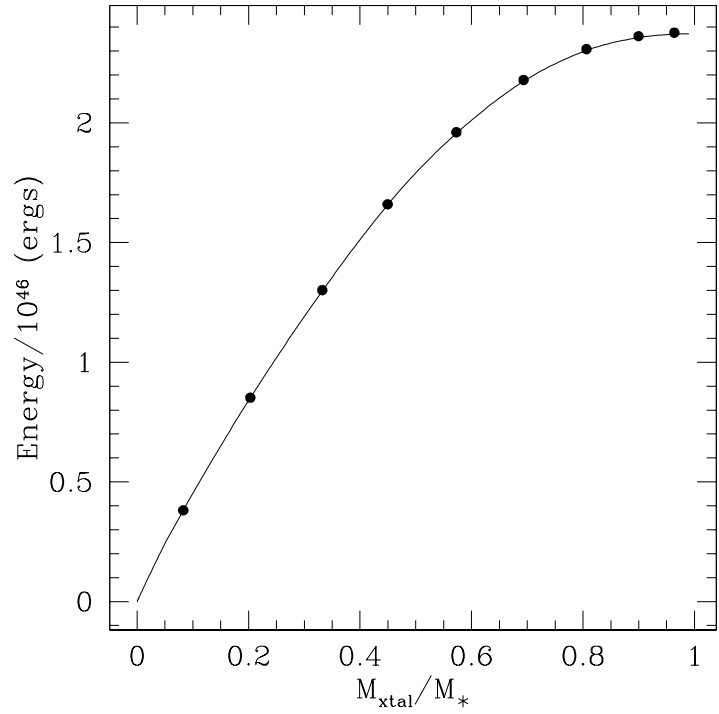,width=3.9in}
\caption{A comparison of the energy released during crystallization
  from a static calculation (line) with that from a self-consistent
  evolutionary calculation (filled dots). The error for the total energy
  released at complete crystallization is less than 0.5\%.
  \label{fig2} }
\end{figure}

Figure \ref{fig2} provides the final justification for our
assumptions.  The filled dots represent the energy released as
computed self-consistently at each evolutionary time step, and the
solid line is the calculated energy released assuming a static
density and temperature profile as described in the above paragraph.
The best agreement is for smaller amounts of crystallization, since
these models differ the least from the initial static model. Even near
complete crystallization, however, the difference between the two
values is less than 0.5\%, justifying our assumptions.  Computationally,
it is very convenient to compute the energy release just once at
the outset and then interpolate using the present value of the
crystallized mass fraction. This allows WDEC to avoid doing a
calculation of the energy release for each iteration of each model,
which would significantly affect the speed of the calculations.

Because all our calculations are done on evolutionary timescales, we do
not have any information about the actual dissipative processes which
are responsible for depositing the energy of phase separation locally.
Indeed, without an accurate hydrodynamic model of the mixing process,
this is not possible.  Fortunately, it is more important to know the
total energy released rather than exactly how this energy is deposited
within the white dwarf model.  This is because the core has a very high
thermal conductivity, which tends to smooth out the temperature
distribution. Thus, wherever the energy is initially deposited, it will
soon be shared throughout the core; indeed, an isothermal core was an
assumption of the original \cite{Mestel52} theory (1952), and it is
still a very accurate description of the physics in the interiors of
white dwarfs (e.g., \cite{Garcia-Berro96} 1996; \cite{Segretain94} 1994).
We therefore choose $\epsilon\subs{ps}$ such that the local temperature
is increased by the same fractional amount throughout the core, i.e.,
$\frac{\delta T}{T} =$ const., while we simultaneously require that the
total energy deposited in this way is equal to the energy released due
to phase separation in a given timestep.  This is somewhat analogous to
the analytical approach outlined by \cite{Isern97} (1997), although we
developed our approach for ease of numerical implementation.

There is one final adjustment which we make to the value of
$\epsilon\subs{ps}$ as calculated above. It is due to the fact that
WDEC calculates models quasi-statically, so that $\epsilon\subs{ps}$ is
assumed to have been constant during the last time step taken, when in
fact it may have changed by a substantial amount. Put another way, the
value of $\epsilon\subs{ps}$ which WDEC calculates should be associated
with the average luminosity of the present and previous timesteps, not
just the current luminosity. Thus, WDEC is implicitly calculating a
delay based upon
\begin{equation}
t_d = \sum_i \frac{\Delta E_i}{L_i}
\label{td3}
\end{equation}
instead of the expression in equation \ref{td2}. We can remedy this
situation by an appropriate rescaling of $\epsilon\subs{ps}$. If
we rescale $\Delta E_i$, and hence $\epsilon\subs{ps}$, by $2
L_i/(L_i+L_{i-1})$, then equation \ref{td3} is transformed into equation
\ref{td2}, and we recover the correct age delay due to crystallization
when implemented in the evolution code. In the limit that our timesteps
are very small, the above prescription is not necessary, but such small
timesteps would be computationally inconvenient, both from a cpu-time
standpoint and from a numerical convergence standpoint.

\subsection{Consistency Checks}
\label{consistency}
\begin{figure}
\epsfig{file=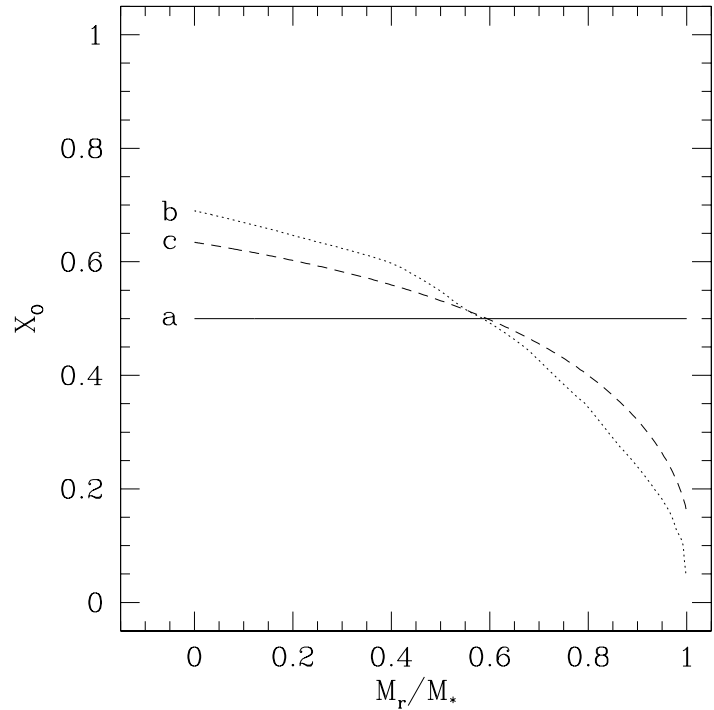,width=3.9in}
\caption{Before crystallization has occurred, we have assumed a 50:50
  C/O mixture, as shown by the solid line (a). After crystallization
  is complete, the oxygen profile is given by the dotted line (b) if the
  phase diagram of \cite{Segretain93} (1993) is used, and by the dashed line
  (c) if the phase diagram of Ichimaru \etal\ (1988) is used. We have
  assumed complete mixing of the remaining fluid layers at each stage
  of crystallization. It is the redistribution of matter from the
  initial to the final profile which results in a net decrease in
  the overall binding energy of the configuration. This model is
  for a 0.6 $M_{\odot}$ white dwarf.
  \label{fig3} }
\end{figure}

We use three different initial C/O profiles in our analysis.  In Figure
\ref{fig3} we show the oxygen composition in the core both before
(dotted line) and after (solid line) crystallization has taken place. We
have taken a homogeneous 50:50 C/O initial distribution and assumed complete
mixing of the overlying fluid layers as crystallization takes place. This
should place an upper limit on the effect which phase separation can have
on any particular model.  We note that the composition profile after
crystallization assuming the \cite{Segretain93} (1993) phase diagram
agrees well with that given in \cite{Chabrier93} (1993).

\begin{figure}
\epsfig{file=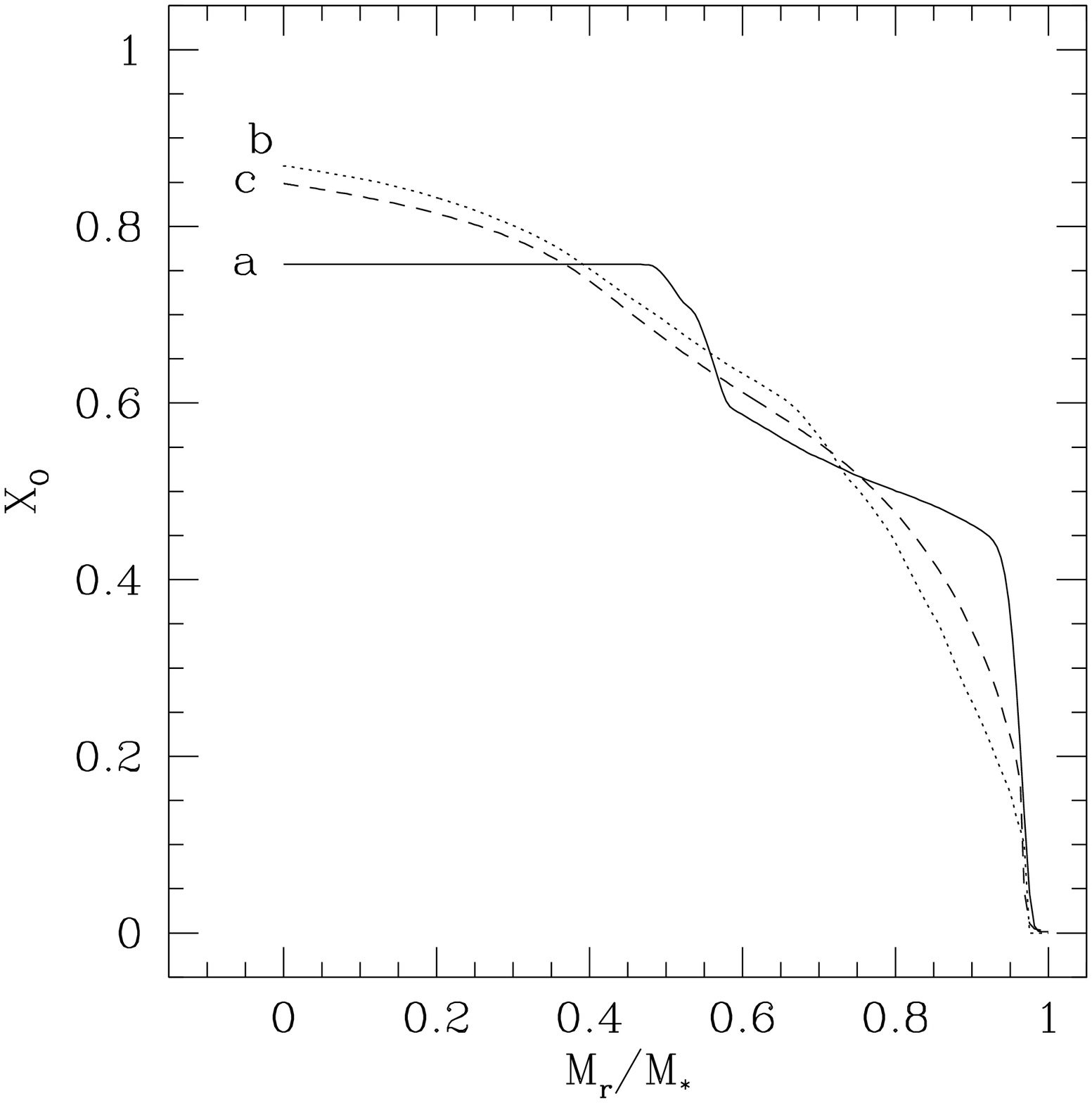,width=3.9in}
\caption{The same as Figure \protect\ref{fig3}, except that
  the initial C/O profile (a) is that computed by \cite{Salaris97}
  (1997) for a 0.61 $M_{\odot}$ white dwarf model. Curves (b) and (c)
  are the final profiles assuming the \cite{Segretain93} (1993) and
  Ichimaru \etal\ (1988) phase diagrams, respectively.  Note that the
  oxygen mass-fraction at the very center increases by only about 15\%
  during crystallization in this case, as compared with a 40\% increase
  for the central value in Figure \protect\ref{fig3}.  Thus, less
  energy is liberated.
  \label{fig4} }
\end{figure}

Figure \ref{fig4} shows a different initial oxygen profile which is
computed in \cite{Salaris97} (1997) for a 0.61 $M_{\odot}$ white dwarf
model. This profile was obtained by considering nuclear reaction processes
in the white dwarf progenitor.  Here we use a modified algorithm for
mixing which reduces to the ``complete mixing'' algorithm when applied to
an initially flat distribution.  When a shell crystallizes, we check to
see if the enhanced carbon content of the innermost fluid shell now has
more carbon than the shell overlying it.  If it does, then we mix the
two shells and perform the same comparison with the next shell farther
out, mixing all three shells if necessary. In this way, we move outward
through the fluid until further mixing no longer decreases the carbon
content of the fluid between this point and the crystallization boundary.
This is physically reasonable, since carbon is, in this sense, ``lighter''
than oxygen, so these layers should be mixed by a convective instability.

For completeness, we use a third profile taken from \cite{Wood90} (1990)
and \cite{Wood95} (1995). It is designed to be representative of C/O profiles
calculated in \cite{Mazzitelli86} (1986) and \cite{D'Antona89}
(1989), who also consider nuclear reaction rates.  
Algebraically, it is given by
\begin{equation}
  X\subs{ox} = \left\{ \begin{array}{ll}
    0.75 & 0.0 \le q \le 0.5 \\
    0.75-1.875 (q -0.5) & 0.5 < q \le 0.9 \\
    0.0 & 0.9 < q \le 1.0
    \end{array}
    \right.
\label{stratified}
\end{equation}
where $q = M\subs{r}/M_{\star}$ and $X\subs{c} = 1 - X\subs{ox}$.

Our treatment of the mixing process provides an upper bound
for the efficiency of this process. If we were to perform a more
self-consistent calculation, we would compute the \bvfreq\ for a given
chemical composition profile in the model and mix those layers which
were convectively unstable {\em and} whose computed timescales for
mixing were shorter than the individual timesteps in our evolutionary
calculations.  An analytical approach to this more detailed problem is
given in \cite{Isern97} (1997) and \cite{Mochkovitch83} (1983). Here
we merely note that a typical value of $|N^2|$ for a Rayleigh-Taylor
unstable region in the cores of our models is $\sim 10^{-4}$, yielding
a timescale for the mixing instability of $\frac{1}{|N|} \sim 10^2$ s,
which is clearly shorter than the relevant timescales for evolution.

\section{A Simple Test Problem}

As a check of the standard approach to treating phase separation, we
performed a simplified treatment of that given in \cite{Xu92} (1992), in
which they calculate the change in binding energy of a zero-temperature
C/Fe white dwarf which undergoes phase separation.  In order to do this,
we have written a separate code which implements the equations for a
zero-temperature degenerate electron gas. Our approach is simpler in that
we do not include Coulomb effects in our EOS calculations, so our approach
is essentially pure Chandrasekhar theory (\cite{Chandrasekhar39} 1939).
We do include, however, relativistic effects, which \cite{Xu92} are unable
to treat. In testing this approach to phase separation, we compute the
energy released due to phase separation in two different ways.  First,
we directly compute the global change in the binding energy.  Second, we
use the expression for the local energy release and integrate this over
the mass of the model, as given in equation \ref{pscalc}. To further
simplify things, we have taken the initial state to be one in which the
distribution of Fe and C is uniform throughout the model, and we have
taken the final state to be a pure Fe core surrounded by a pure C mantle.

\begin{figure}
\epsfig{file=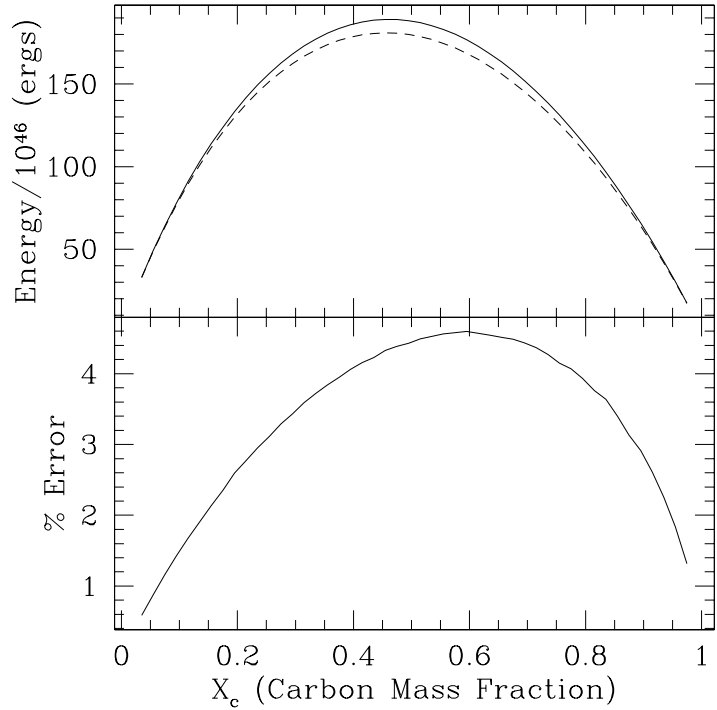,width=3.9in}
\caption{The upper panel shows the energy released due to phase separation
as a function of $X\subs{c}$, the carbon mass-fraction. The model is
a C/Fe mixture computed assuming pure Chandrasekhar theory. The solid
line comes from a direct calculation of the change in binding energy,
and the dotted line is obtained from the application of equation
\protect\ref{pscalc}. The lower panel shows the percent error between
these two methods. The total mass of the model is set to 0.66546
$M_{\odot}$, as in \protect\cite{Xu92} (1992).
\label{fig5} }
\end{figure}

Figure \ref{fig5} shows the results for differing initial fractions
of C and Fe. For instance, for an initial 50:50 C/Fe distribution we
calculate an energy release of about $1.9 \cdot 10^{48}$ ergs, with
less than a 5\% relative error between the two methods. Even this small
amount of error decreases as we approach a pure Fe or C initial state.
This is because the density and composition changes before and after
phase separation are now smaller, which makes the local calculation more
accurate. For instance, if the model is 99\% C uniformly distributed
initially, then after phase separation most of it (99\% in fact) is
pure C. The contrast between 99\% and 100\% is small enough that the
local density and composition changes are also small ($\delta X\subs{C}
\lesssim $ 1\%, $\frac{\delta \rho}{\rho} \lesssim$ 0.1\%), which means
that the approximation involved in making the infinitesimal variations
in equation \ref{pscalc} finite is more accurate.  We note that it is
possible to perform such a simplified treatment for a C/Fe white dwarf
model and still obtain meaningful results, while for a C/O model it would
not be possible. This is because $\mu\subs{e} = 2.0$ for both C and O,
while $\mu\subs{e} = 2.15$ for Fe. Thus, ignoring Coulomb effects,
C and O have identical equations of state, while C and Fe are still
nontrivially different in this approximation.

The results of this test problem (Figure \ref{fig5}) convince us
that by applying equation \ref{pscalc}, we are correctly
calculating the change in the binding energy of the configuration, and
thus the amount of thermal energy which has been liberated from structural 
sources. This shows us that the overall approach to this problem which
we use here, and which has been used in the past, is sound and accurately
describes the physics of phase separation.

\section{Results}
\subsection{0.6 $M_{\odot}$ White Dwarf Models}
\label{fiducial}

In Table \ref{tab1} we give an evolutionary listing of our fiducial
sequence (other sequences are available from the author upon request).
This sequence is more than just a convenient reference model
for the rest of our calculations. Given the observed peak of the
masses of isolated white dwarfs in the vicinity of 0.6 $M_{\odot}$
(\cite{Weidemann83} 1983; \cite{Weidemann89} 1989; \cite{Bergeron95}
1995; \cite{Lamontagne97} 1997), this model will be the most useful
in our comparisons with the white dwarf population as a whole. For the
surface layer masses, we have taken $M_{\rm He}/M_{\star}=10^{-2}$ and
$M_{\rm H}/M_{\star}=10^{-4}$ as in \cite{Wood95} (1995). We explore
the effect of different surface layer masses later in this section.

\begin{deluxetable}{ccccccc}
\footnotesize
\tablewidth{33pc}
\tablecaption{Fiducial White Dwarf Cooling Sequence without Phase Separation
\protect\label{tab1}
}
\tablehead{\colhead{$\log L/L_{\odot}$} & \colhead{$\log$ Age (yr)}
&  \colhead{$\log T_{c}$} & \colhead{$\log T_{\rm eff}$} & 
\colhead{$\log R_{\star}$} & 
\colhead{$\log L_{\nu}/L_{\odot}$} & 
\colhead{$M_{\rm xtal}/M_{\star}$}}
\startdata
 1.0000 & 5.917 & 7.950 & 4.869 & 9.131 &  1.298 & 0.000 \\ 
 0.6000 & 6.158 & 7.891 & 4.793 & 9.084 &  0.946 & 0.000 \\ 
 0.2000 & 6.376 & 7.843 & 4.710 & 9.049 &  0.581 & 0.000 \\ 
-0.2000 & 6.592 & 7.798 & 4.623 & 9.024 &  0.193 & 0.000 \\ 
-0.6000 & 6.855 & 7.737 & 4.533 & 9.004 & -0.230 & 0.000 \\ 
-1.0000 & 7.204 & 7.660 & 4.440 & 8.989 & -0.886 & 0.000 \\ 
-1.2000 & 7.429 & 7.604 & 4.394 & 8.982 & -1.356 & 0.000 \\ 
-1.4000 & 7.674 & 7.531 & 4.346 & 8.976 & -1.956 & 0.000 \\ 
-1.6000 & 7.903 & 7.448 & 4.299 & 8.971 & -2.670 & 0.000 \\ 
-1.8000 & 8.097 & 7.360 & 4.251 & 8.967 & -3.449 & 0.000 \\ 
-2.0000 & 8.264 & 7.273 & 4.203 & 8.963 & -4.248 & 0.000 \\ 
-2.2000 & 8.413 & 7.187 & 4.155 & 8.959 & -5.042 & 0.000 \\ 
-2.4000 & 8.550 & 7.103 & 4.106 & 8.956 & -5.804 & 0.000 \\ 
-2.6000 & 8.679 & 7.021 & 4.058 & 8.953 & -7.279 & 0.000 \\ 
-2.8000 & 8.805 & 6.940 & 4.009 & 8.951 & < -10.00 & 0.000 \\ 
-3.0000 & 8.930 & 6.860 & 3.961 & 8.948 & < -10.00 & 0.000 \\ 
-3.2000 & 9.055 & 6.778 & 3.912 & 8.946 & < -10.00 & 0.000 \\ 
-3.4000 & 9.182 & 6.694 & 3.863 & 8.944 & < -10.00 & 0.000 \\ 
-3.6000 & 9.317 & 6.613 & 3.814 & 8.942 & < -10.00 & 0.059 \\ 
-3.8000 & 9.497 & 6.518 & 3.765 & 8.939 & < -10.00 & 0.379 \\ 
-4.0000 & 9.704 & 6.350 & 3.717 & 8.935 & < -10.00 & 0.804 \\ 
-4.2000 & 9.825 & 6.176 & 3.668 & 8.933 & < -10.00 & 0.961 \\ 
-4.4000 & 9.900 & 6.022 & 3.618 & 8.933 & < -10.00 & 0.989 \\ 
-4.6000 & 9.953 & 5.891 & 3.569 & 8.932 & < -10.00 & 0.990 \\ 
\enddata
\end{deluxetable}

\begin{figure}
\epsfig{file=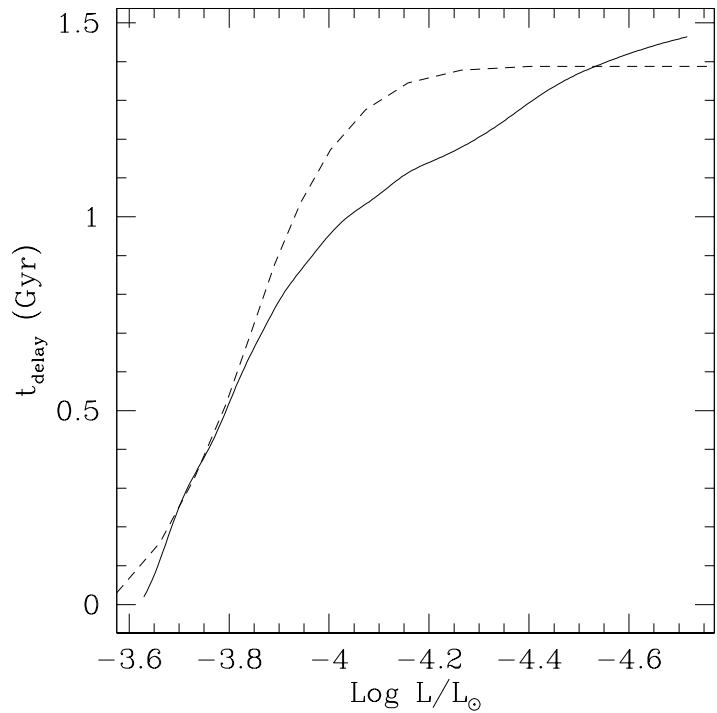,width=3.9in}
\caption{The solid line is a self-consistent calculation of the age
difference between two 0.6 $M_{\odot}$ white dwarf evolutionary sequences
with $Z=0.0$, one of which is undergoing phase separation. The dotted line
is the result of applying equation \protect\ref{td2} to the evolutionary
sequence undergoing phase separation, which yields an asymptotic value
for the age delay of $\sim$ 1.4 Gyr.  At complete crystallization ($\log
L/L_{\odot} \sim -4.6$), the value given by the direct evolutionary
calculation is within 5\% of this, indicating that the basic physics
which is operating is well-described by equation \protect\ref{td1}.
\label{fig6} }
\end{figure}

We now wish to consider the effect of phase separation on actual
evolutionary sequences.  We compute this effect in two different ways.
Using the first method, we compare sequences in which the physics of phase
separation is included with those in which it is not. Taking two such
sequences, we first perform a spline fit for each sequence's age over
a fixed luminosity grid, and then we calculate the difference in ages
at each luminosity.  The results are given by the solid line in Figure
\ref{fig6}, and indicate an age delay at complete crystallization of
about 1.5 Gyr; this is for an initially homogeneous 50:50 C/O profile
with an assumed metallicity of $Z=0.000$ in the opacities. We note that
all the age delays computed here are for complete crystallization of our
models, and hence represent upper limits to the possible age delays.
Operationally, our models are almost completely crystallized near the
observed luminosity turndown at $\log L_{\star}/L_{\odot} \sim -4.5$,
so that there is at most a 1\% change in our calculated age delays if
we only consider models which have not yet cooled beyond this point.

The other method involves applying equation \ref{td2} to a sequence
undergoing phase separation, which is shown by the dotted line in Figure
\ref{fig6}.  This yields an asymptotic value for the age delay of 1.38
Gyr, which is within 5\% of the age difference computed with the first
method.  This result indicates that the basic physics which is operating
is well-described by equation \ref{td2}, i.e., the energy being released
by phase separation is mostly being radiated in a given timestep.  For the
remainder of the results quoted here, the age delays have been calculated
using this second method (equation \ref{td2}), since this proves to be more
accurate for the cases involving smaller energy releases and age delays.

We now study the effect of the initial composition profile on the age
delays. We use three different profiles: one which is a homogeneous 50:50
mix (Figure \ref{fig3}), one calculated by \cite{Salaris97} (1997)
(Figure \ref{fig4}), and one given by equation \ref{stratified}.
Our results are summarized in Table \ref{tab2}, where the columns
labeled SC and IIO indicate that we have used the phase diagrams of
\cite{Segretain93} (1993) and Ichimaru \etal\ (1988), respectively.
Near the centers of these models, we found that the initial/final oxygen
mass-fraction changed by only about 15\% in the initially stratified case
in Figure \ref{fig4}, as compared to 40\% in the homogeneous case
in Figure \ref{fig3}.  Because less matter is being redistributed
in the initially stratified case, we would expect less energy to be
released as a result. Using the phase diagram of \cite{Segretain93}
(1993) applied to a 0.6 $M_{\odot}$ white dwarf model, we find that in
the homogeneous case $2.38 \cdot 10^{46}$ ergs are released whereas
in the initially stratified case in Figure \ref{fig4} only $1.03
\cdot 10^{46}$ ergs are released. These energies result in age delays
of 1.38 Gyr and 0.62 Gyr, respectively. Thus, the initial composition
profile has a large effect on the calculated age delays. In addition,
the Ichimaru \etal\ (1988) phase diagram produces smaller composition
changes and hence smaller values, reducing the \cite{Segretain93} age
delays by approximately one-third.

\begin{planotable}{lcc}
\footnotesize
\tablewidth{25pc}
\tablecaption{Age Delays for $0.6 M_{\odot}$ Models \label{tab2}}
\tablehead{\colhead{Initial Profile} & \multicolumn{2}{c}{Delay (Gyr)}\nl
 \colhead{} & \colhead{SC} & \colhead{IIO}}
\startdata
50:50 homogeneous & 1.38 & 0.99 \nl 
stratified (\cite{Salaris97} 1997) & 0.62 & 0.39 \nl
stratified (\cite{Wood95} 1995) & 0.30 & 0.20 \nl
\enddata
\end{planotable}

We now consider the effect of a nonzero metallicity in the opacity
tables. The effect of varying the metallicity from $Z=0.000$ to
$Z=0.001$ results in a change of less than 0.016\% in the energies
released, and is barely detectable numerically. The main effect of
changing the metallicity is to affect the luminosity range at which the
phase separation energy is released, which in turn affects the age
delay, $t\subs{delay}$. For both the homogeneous and stratified case,
the average luminosity during crystallization changes by less than 3\%
as $Z$ is varied from 0.000 to 0.001, and hence $t\subs{delay}$ also
changes by less than 3\%. Thus, the age delay is essentially
insensitive to the metallicity assumed for the opacities.

Finally, we summarize the effect of different surface layer masses in
Table~\ref{tab3}. For $M_{\rm He}/M_{\star}=10^{-3}$ and $M_{\rm
H}/M_{\star}=10^{-5}$ (comp1), we find maximum age delays of 1.45 Gyr,
and for $M_{\rm He}/M_{\star}=10^{-4}$ and $M_{\rm H}/M_{\star}=10^{-6}$
(comp2), our maximum calculated age delay is 1.56 Gyr. These values
represent increases of 5\% and 13\%, respectively, over the age delays
calculated in our fiducial model.  For clarity, we note that these
calculations are for the age differences introduced by phase separation
alone at these new surface layer masses; the white dwarf ages themselves
change significantly with He layer mass, which produces a decrease in
the calculated ages (without including phase separation) of $\sim 0.75$
Gyr for each order of magnitude increase in $M_{\rm He}$. Again, we
find that varying the metallicity in the opacities has a small effect
on these numbers, at only the 1\% level.

\begin{planotable}{lcccc}
\footnotesize
\tablewidth{30pc}
\tablecaption{Age Delays for $0.6 M_{\odot}$ Models with Different Surface
  Layer Masses\label{tab3}}
\tablehead{\colhead{Initial Profile} & \multicolumn{4}{c}{Delay (Gyr)}\nl
 \colhead{} & \multicolumn{2}{c}{comp1} & \multicolumn{2}{c}{comp2}\nl
 \colhead{} & \colhead{SC} & \colhead{IIO} & \colhead{SC} & \colhead{IIO} }
\startdata
50:50 homogeneous & 1.45 & 1.04 & 1.56 & 1.12 \nl 
stratified (\cite{Salaris97} 1997) & 0.66 & 0.42 & 0.71 & 0.44  \nl
stratified (\cite{Wood95} 1995) & 0.32 & 0.21 & 0.34 & 0.23 \nl
\enddata
\end{planotable}

\subsection{The Mass Dependence}
\label{masses}

\begin{figure}
\epsfig{file=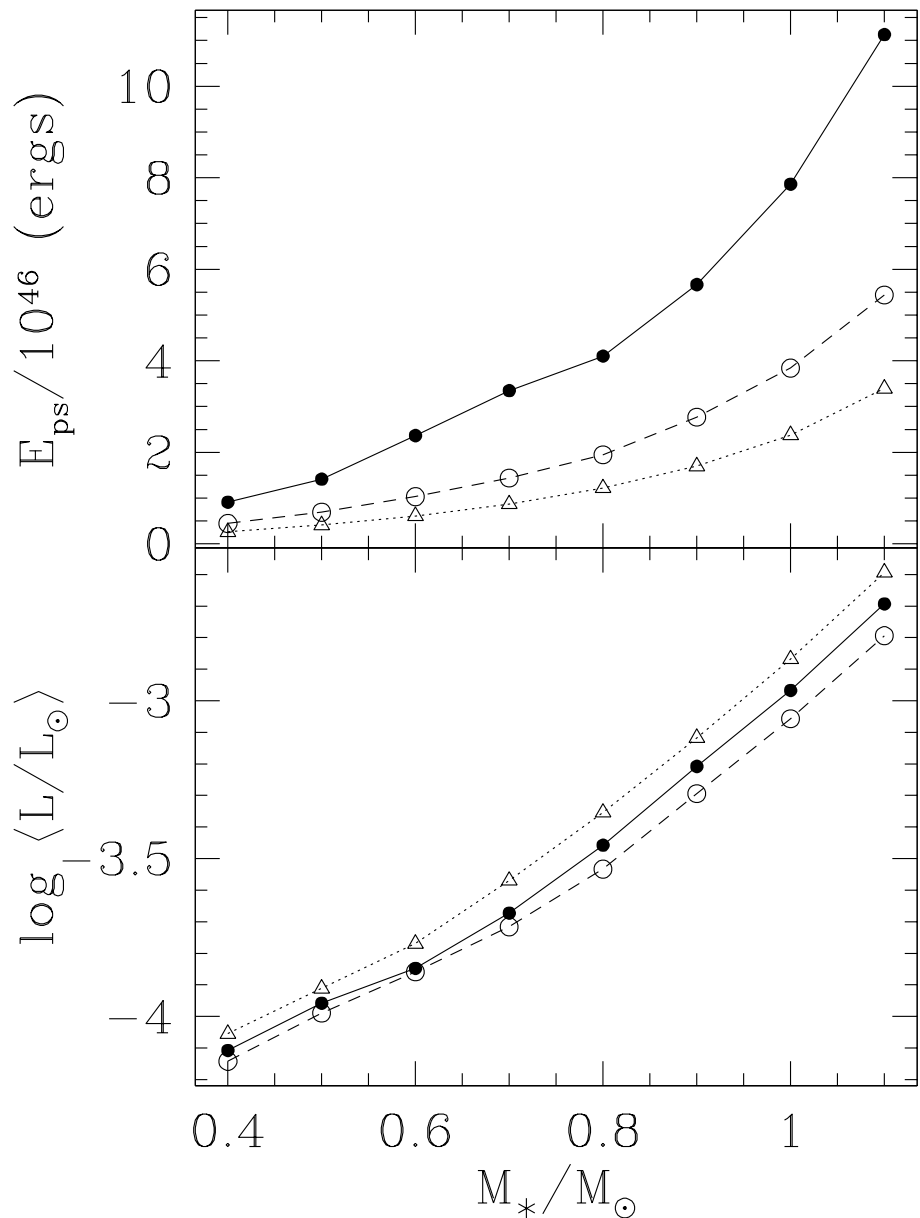,width=5.0in}
\caption{Phase separation energy and average luminosity as a function
  of mass, using the Segretain \& Chabrier phase diagram. The solid
  curves are for initially homogeneous 50:50 C/O mixtures, the dashed
  curves are for the stratified C/O profile of Figure \protect\ref{fig4}
  (\cite{Salaris97} 1997), and the dotted curves are for the stratified
  C/O profile of equation \protect\ref{stratified} (\cite{Wood95} 1995).
  The upper panel shows the total phase separation energy released as a
  function of total stellar mass, and the lower panel shows the average
  luminosity during the crystallization process, also as a function of
  total stellar mass.
  \label{fig7}
  }
\end{figure}

The mass of the white dwarf model affects the process of phase separation
in two main ways, as is illustrated in Figure \ref{fig7}. First, a more
massive white dwarf has a higher gravity so that more energy is released
by the subsequent rearrangement of matter.  Second, the luminosity
at which crystallization occurs is higher for a more massive white
dwarf, which tends to lessen the age-delay for a given energy release.
For example, even though the total energy released in a 1.2 $M_{\odot}$
model increases by a factor of $\sim 10$, the average luminosity increases
by a factor of $\sim 30$, and hence there is a net decrease in the time
delay relative to the 0.6 $M_{\odot}$ sequence.

\begin{figure}
\epsfig{file=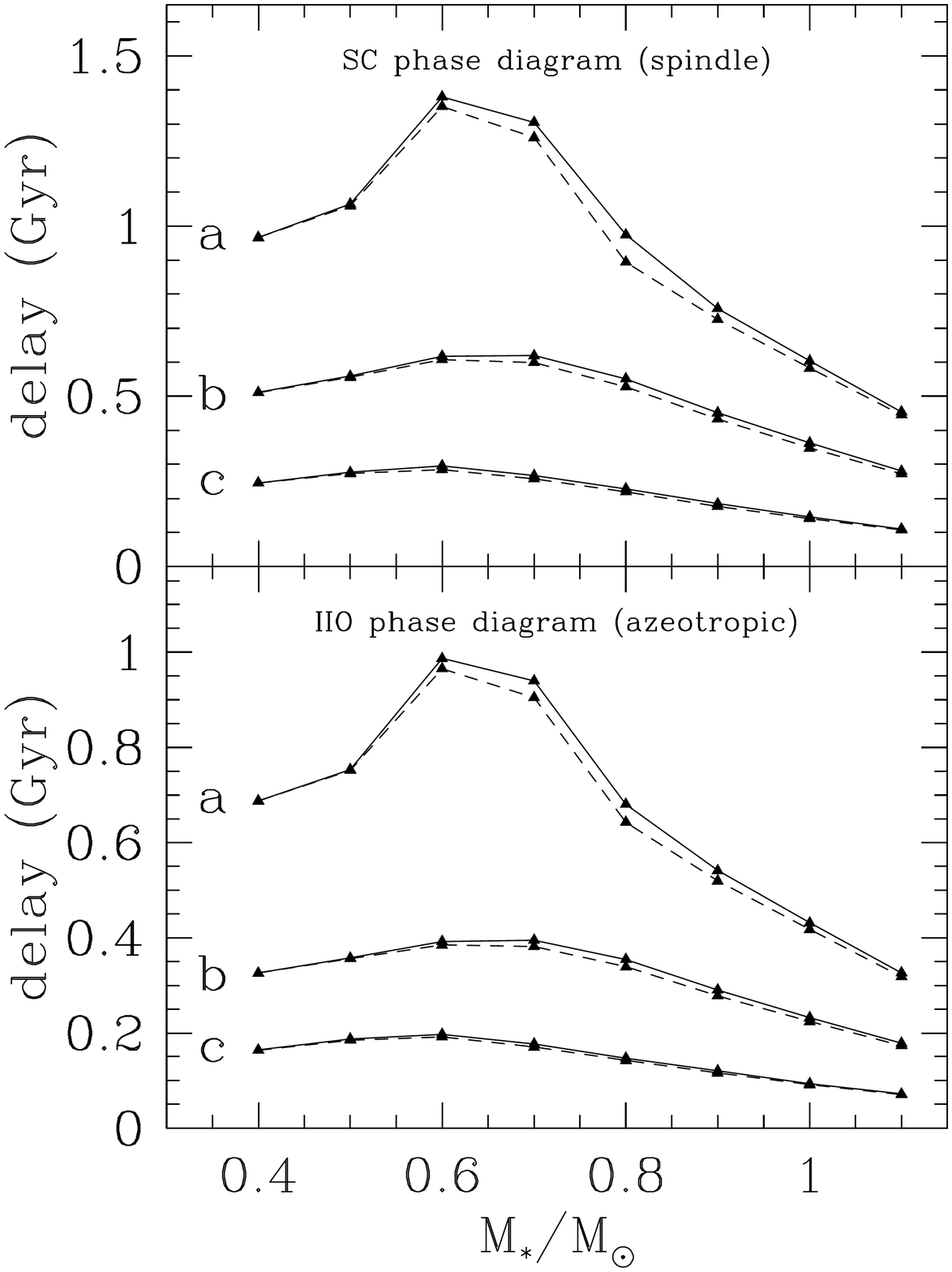,width=5.0in}
\caption{Age delay due to phase separation during crystallization as a
function of total mass of the white dwarf model. Curve (a) corresponds
to a 50:50 homogeneous initial C/O profile, while curves (b) and
(c) are the initial profiles shown by the solid lines in Figures
\protect\ref{fig4} and \protect\ref{fig3}, respectively.
The solid lines are for zero metallicity opacities and the dashed lines
are for $Z=0.001$, which shows that our result has little metallicity
dependence.  All models have $M_{\rm He}/M_{\star}=10^{-2}$ and $M_{\rm
H}/M_{\star}=10^{-4}$.
\label{fig8} }
\end{figure}

The competition of these two effects suggests that there may be a mass
for which there is a maximum age-delay, for a fixed composition
profile.  This is indeed the case, as is demonstrated in Figure
\ref{fig8}. We find that the 0.6 $M_{\odot}$ white dwarf models have the
maximum age-delays for a given composition profile (this was also found
by \cite{Segretain94} 1994). The calculated age delay is only weakly
dependent upon the metallicity, as can be seen from the small
difference between the solid and dashed curves.  It is strongly
dependent upon the initial profile, however, which can decrease the
energy release, and hence the age-delays, by a factor of three or more,
as is shown in Figure \ref{fig8}.

From the preceding calculations we find that the two most important
factors influencing the magnitude of the age delays introduced by the
physics of phase separation are the {\em mass} of the white dwarf
model and its {\em initial} C/O profile. Because the mass range of
observed white dwarfs is strongly peaked around $0.6 M_{\odot}$ (e.g.,
\cite{Lamontagne97} 1997), we find that the age delay we calculate is
near the maximum possible with respect to this parameter. In terms of
the initial C/O profile, however, the situation is reversed. For a $0.61
M_{\odot}$ white dwarf model, the profile calculated by \cite{Salaris97}
(1997) reduces the age delay by a factor of $\sim 2$ from the 50:50
homogeneous case. Using the profile of \cite{Wood95} (1995), which is
based on results from \cite{Mazzitelli86} (1986) and \cite{D'Antona89}
(1989), the reduction factor is $\sim 5$.

If we take as our best guess the initial profile of \cite{Salaris97}
(1997), assume a $0.6 M_{\odot}$ white dwarf model with $M_{\rm
He}/M_{\star}=10^{-2}$ and $M_{\rm H}/M_{\star}=10^{-4}$, and use 
the \cite{Segretain93} (1993) phase diagram,
then we obtain an age delay of $\sim$ 0.6 Gyr.

\section{Conclusions}

We find a maximum age delay of $\sim 1.5$ Gyr due to phase separation for
our fiducial white dwarf model ($M_{\star}=0.6 M_{\odot}$) and a best
guess age delay of $\sim$ 0.6 Gyr. \cite{Salaris97} (1997) have recently
calculated a value of $\sim$ 1 Gyr, using the evolutionary models of
\cite{Wood89} (1989). If we scale their value to our present models
(assuming an average luminosity during crystallization for their models of
$\log L/L_{\odot} \simeq -4.1$), then we obtain $0.75$ Gyr, which is in
basic agreement with our estimate of 0.62 Gyr.  The differences in these
models are mainly due to the different surface layer masses adopted;
more recent asteroseismological analyses of the class of DA's suggests
that the appropriate surface layer masses are $M_{\rm He}/M_{\star}\sim
10^{-2}$ and $M_{\rm H}/M_{\star}\sim 10^{-4}$ (\cite{Clemens93} 1993,
\nocite{Clemens95} 1995), and these are the values which we have assumed.

The most important factors influencing the size of the calculated
age delay are the total stellar mass and the initial composition
profile.  We find the largest age delays occur in models with masses
of $\sim 0.6 M_{\odot}$, near the peak in the observed white dwarf
mass distribution. The best current theoretical initial C/O profile
produces models with smaller age delays, of $\sim 0.6$ Gyr. In addition,
if we use the phase diagram of Ichimaru \etal\ (1988) instead of the
\cite{Segretain93} (1993) phase diagram, then our age delays are reduced
by about one-third. We note that the prescription which we have adopted
for the mixing during crystallization provides an upper bound for the
efficiency of this process, and hence a maximum for the age delay. More
realistic treatments of the mixing process may reduce the age delay.
We find that varying the opacities (via the metallicity) and varying
the surface layer masses has only a small effect ($\lesssim$ 10\%)
on the calculated age delays.

Our calculations do not take into account the possible age delays
introduced by the phase separation of heavier trace-element species
such as $^{22}$Ne, which may produce significant age delays of 2--3 Gyr
(\cite{Segretain94} 1994; \cite{Hernanz94} 1994).  These species would
arise from the initial abundance of metals in the main-sequence stars
which later evolved into white dwarfs.  This effect may only be
important for Population I stars, however, and would not therefore
affect the calculated ages of the cool white dwarfs which populate the
turndown in the WDLF, since these white dwarfs were formed very early
in the history of the Galaxy (\cite{Hernanz94} 1994). 

In the context of Galactic evolution, age estimates for the oldest
Galactic globular clusters range from 13--16 Gyr (\cite{Pont98} 1998)
to 11.5 $\pm$ 1.3 Gyr (\cite{Chaboyer98} 1998), and depend on a variety
of parameters. In addition, a 4 to 6 Gyr delay is expected between the
formation of the globular clusters and that of the Galactic thin disk
(e.g., \cite{Burkert92} 1992; \cite{Chiappini97} 1997), while the observed
white dwarf luminosity function gives an age estimate for the thin disk
of $9.5^{+1.1}_{-0.8}$ Gyr (\cite{Oswalt96} 1996), without including the
effect of phase separation. Using the above numbers, we see that phase
separation could add anywhere from 0 to 3 Gyr to the white dwarf ages
and still be consistent with the overall picture of Galaxy formation.
Our calculated maximum value of $\lesssim$ 1.5 Gyr fits within these
bounds, as does our best guess value of $\sim$ 0.6 Gyr.

\section{Acknowledgments}
We would like to thank M. Salaris for providing the oxygen profile
shown in Figure \ref{fig4}, and we would like to thank the referee
G. Chabrier for his valuable comments and suggestions.

This work was supported in part by the National Science Foundation
under grant AST-9315461 (EWK, MHM, and DEW) and grant AST-9217988 (MAW)
and by the NASA Astrophysics Theory Program under grant NAG5-2818 (EWK,
MHM, and DEW) and grant NAG5-3103 (MAW).

%\newpage

\bibliographystyle{../../styles/astro2}
\bibliography{../../styles/index_f}

\begin{thebibliography}{}

\bibitem[{Abrikosov}]{Abrikosov60}
{Abrikosov},~A.~A. 1960, Zh. Eksp. i Teor. Fiz, 39, 1798

\bibitem[{Barrat}, {Hansen}, \& {Mochkovitch}]{Barrat88}
{Barrat},~J.~L., {Hansen},~J.~P., \& {Mochkovitch},~R. 1988, \aap, 199, L15

\bibitem[{Bergeron} {et~al.}]{Bergeron95}
{Bergeron},~P., {Wesemael},~F., {Lamontagne},~R., {Fontaine},~G.,
  {Saffer},~R.~A., \& {Allard},~N.~F. 1995, \apj, 449, 258

\bibitem[{Burkert}, {Truran}, \& {Hensler}]{Burkert92}
{Burkert},~A., {Truran},~J.~W., \& {Hensler},~G. 1992, \apj, 391, 651--658

\bibitem[{Chaboyer} {et~al.}]{Chaboyer98}
{Chaboyer},~B., {Demarque},~P., {Kernan},~P.~J., \& {Krauss},~L.~M. 1998, \apj,
  494, 96

\bibitem[{Chabrier}]{Chabrier98}
{Chabrier},~G. 1998, In Proceedings of IAU Symposium 189 on Fundamental Stellar
  Properties: The Interaction between Observation and Theory, ed. T.~R.
  {Bedding}, A.~J. Booth, \& J.~Davis, volume 189 (Dordrecht: Kluwer), 381

\bibitem[{Chabrier} {et~al.}]{Chabrier93}
{Chabrier},~G., {Segretain},~L., {Hernanz},~M., {Isern},~J., \&
  {Mochkovitch},~R. 1993, In White Dwarfs: Advances in Observation and Theory,
  ed. M.~A. {Barstow} (Dordrecht: Kluwer Academic Publishers), 115

\bibitem[{Chandrasekhar}]{Chandrasekhar39}
{Chandrasekhar},~S.
1939, An Introduction to the Study of Stellar Structure (Chicago: University of
  Chicago Press)

\bibitem[{Chiappini}, {Matteucci}, \& {Gratton}]{Chiappini97}
{Chiappini},~C., {Matteucci},~F., \& {Gratton},~R. 1997, \apj, 477, 765

\bibitem[{Clemens}]{Clemens93}
{Clemens},~J.~C. 1993, PhD thesis, The University of Texas at Austin

\bibitem[{Clemens}]{Clemens95}
{Clemens},~J.~C. 1995, Baltic Astronomy, 4, 142

\bibitem[{D'Antona} \& {Mazzitelli}]{D'Antona78}
{D'Antona},~F. \& {Mazzitelli},~I. 1978, \aap, 66, 453

\bibitem[{D'Antona} \& {Mazzitelli}]{D'Antona89}
{D'Antona},~F. \& {Mazzitelli},~I. 1989, \apj, 347, 934

\bibitem[{DeWitt}, {Slattery}, \& {Chabrier}]{DeWitt96}
{DeWitt},~H., {Slattery},~W., \& {Chabrier},~G. 1996, Physica B, 228, 21

\bibitem[{Garc\'{\i}a-Berro} {et~al.}]{Garcia-Berro96}
{Garc\'{\i}a-Berro},~E., {Hernanz},~M., {Isern},~J., {Chabrier},~G.,
  {Segretain},~L., \& {Mochkovitch},~R. 1996, Astronomy and Astrophysics
  Supplement Series, 117, 13

\bibitem[{Garc\'{\i}a-Berro} {et~al.}]{Garcia-Berro88}
{Garc\'{\i}a-Berro},~E., {Hernanz},~M., {Mochkovitch},~R., \& {Isern},~J. 1988,
  \aap, 193, 141

\bibitem[{Hernanz} {et~al.}]{Hernanz94}
{Hernanz},~M., {Garc\'{\i}a-Berro},~E., {Isern},~J., {Mochkovitch},~R.,
  {Segretain},~L., \& {Chabrier},~G. 1994, \apj, 434, 652

\bibitem[{Ichimaru}, {Iyetomi}, \& {Ogata}]{Ichimaru88}
{Ichimaru},~S., {Iyetomi},~H., \& {Ogata},~S. 1988, \apjl, 334, L17

\bibitem[{Iglesias} \& {Rogers}]{Iglesias93}
{Iglesias},~C.~A. \& {Rogers},~F.~J. 1993, \apj, 412, 752

\bibitem[{Isern} {et~al.}]{Isern97}
{Isern},~J., {Mochkovitch},~R., {Garc\'{\i}a-Berro},~E., \& {Hernanz},~M. 1997,
  \apj, 485, 308

\bibitem[{Kirzhnits}]{Kirzhnits60}
{Kirzhnits},~D.~A. 1960, Soviet Phys.---JETP, 11, 365

\bibitem[{Lamb} \& {Van Horn}]{Lamb75}
{Lamb},~D.~Q. \& {Van Horn},~H.~M. 1975, \apj, 200, 306

\bibitem[{Lamontagne} {et~al.}]{Lamontagne97}
{Lamontagne},~R., {Wesemael},~G., {Fontaine},~G., \& {Demers},~S. 1997, In
  White Dwarfs, Proceedings of the 10th European Workshop on White Dwarfs, ed.
  J.~{Isern}, M.~{Hernanz}, \& E.~{Garc\'{\i}a-Berro} (Dordrecht, Boston,
  London: Kluwer Academic Publishers), 143

\bibitem[{Liebert}, {Dahn}, \& {Monet}]{Liebert88}
{Liebert},~J., {Dahn},~C.~C., \& {Monet},~D.~G. 1988, \apj, 332, 891

\bibitem[{Mazzitelli} \& {D'Antona}]{Mazzitelli86}
{Mazzitelli},~I. \& {D'Antona},~F. 1986, \apj, 308, 706

\bibitem[{Mestel}]{Mestel52}
{Mestel},~L. 1952, \mnras, 112, 583

\bibitem[{Mestel} \& {Ruderman}]{Mestel67}
{Mestel},~L. \& {Ruderman},~M.~A. 1967, \mnras, 136, 27

\bibitem[{Mochkovitch}]{Mochkovitch83}
{Mochkovitch},~R. 1983, \aap, 122, 212

\bibitem[{Ogata} \& {Ichimaru}]{Ogata87}
{Ogata},~S. \& {Ichimaru},~S. 1987, Phys. Rev. A, 42, 5451

\bibitem[{Oswalt} {et~al.}]{Oswalt96}
{Oswalt},~T.~D., {Smith},~J.~A., {Wood},~M.~A., \& {Hintzen},~P. 1996, \nat,
  382, 692

\bibitem[{Pont} {et~al.}]{Pont98}
{Pont},~F., {Mayor},~M., {Turon},~C., \& {Vandenberg},~D.~A. 1998, \aap, 329,
  87

\bibitem[{Salaris} {et~al.}]{Salaris97}
{Salaris},~M., {Dominguez},~I., {Garc\'{\i}a-Berro},~E., {Hernanz},~M.,
  {Isern},~J., \& {Mochkovitch},~R. 1997, \apj, 486, 413

\bibitem[{Salpeter}]{Salpeter61}
{Salpeter},~E. 1961, \apj, 134, 669

\bibitem[{Schmidt}]{Schmidt59}
{Schmidt},~M. 1959, \apj, 129, 243

\bibitem[{Schwarzschild}]{Schwarzschild58}
{Schwarzschild},~K.
1958, Structure and Evolution of the Stars (Princeton: Princeton University
  Press)

\bibitem[{Segretain} \& {Chabrier}]{Segretain93}
{Segretain},~L. \& {Chabrier},~G. 1993, \aap, 271, L13

\bibitem[{Segretain} {et~al.}]{Segretain94}
{Segretain},~L., {Chabrier},~G., {Hernanz},~M., {Garc\'{\i}a-Berro},~E.,
  {Isern},~J., \& {Mochkovitch},~R. 1994, \apj, 434, 641

\bibitem[{Stevenson}]{Stevenson77}
{Stevenson},~D.~J. 1977, Proc. Ast. Soc. Australia, 3, 167

\bibitem[{Stevenson}]{Stevenson80}
{Stevenson},~D.~J. 1980, J. Phys. Suppl., 41, C2--61

\bibitem[{Van Horn}]{VanHorn68}
{Van Horn},~H.~M. 1968, \apj, 151, 227

\bibitem[{Weidemann} \& {Koester}]{Weidemann83}
{Weidemann},~V. \& {Koester},~D. 1983, \aap, 121, 77

\bibitem[{Weidemann} \& {Yuan}]{Weidemann89}
{Weidemann},~V. \& {Yuan},~J.~W. 1989, In White dwarfs; Proceedings of IAU
  Colloquium 114th, Hanover, NH, ed. G.~{Wegner} (Berlin, New York:
  Springer-Verlag), 1

\bibitem[{Winget} {et~al.}]{Winget87}
{Winget},~D.~E., {Hansen},~C.~J., {Liebert},~J., {Van Horn},~H.~M.,
  {Fontaine},~G., {Nather},~R.~E., {Kepler},~S.~O., \& {Lamb},~D.~Q. 1987,
  \apjl, 315, L77

\bibitem[{Wood}]{Wood93}
{Wood},~M. 1993, Bull. American Astron. Soc., 183, 5002

\bibitem[{Wood}]{Wood95}
{Wood},~M. 1995, In White Dwarfs: Proceedings of the 9th European Workshop on
  White Dwarfs, ed. D.~{Koester} \& K.~{Werner} (Berlin, Heidelberg:
  Springer-Verlag), 41

\bibitem[{Wood}]{Wood90}
{Wood},~M.~A. 1990, PhD thesis, The University of Texas at Austin

\bibitem[{Wood}]{Wood92}
{Wood},~M.~A. 1992, \apj, 386, 539--561

\bibitem[{Wood} \& {Oswalt}]{Wood98}
{Wood},~M.~A. \& {Oswalt},~T.~D. 1998, \apj, 497, 870

\bibitem[{Wood} \& {Winget}]{Wood89}
{Wood},~M.~A. \& {Winget},~D.~E. 1989, In White dwarfs; Proceedings of IAU
  Colloquium 114th, Hanover, NH, ed. G.~{Wegner} (Berlin, New York:
  Springer-Verlag), 282

\bibitem[{Xu} \& {Van Horn}]{Xu92}
{Xu},~Z.~W. \& {Van Horn},~H.~M. 1992, \apj, 387, 662

\end{thebibliography}

\end{document}